\documentclass[onecolumn]{pasj01}
\usepackage{comment}
\usepackage{url}
\usepackage{lineno}
\def\nh{NH$_3$} 

\def\kms{km s$^{-1}$}

\draft 
\Received{2021/11/30}
\Accepted{2022/02/03}
\Published{$\langle$publication date$\rangle$}

\begin{document}
\title{{Ammonia mapping observations toward the Galactic massive star-forming region Sh 2-255 and Sh 2-257} }

\author{Mikito \textsc{Kohno}\altaffilmark{1}$^{*}$}%
\author{Toshihiro \textsc{Omodaka}\altaffilmark{2}}
\author{Toshihiro \textsc{Handa}\altaffilmark{2,3}}
\author{James O. \textsc{Chibueze}\altaffilmark{4,5}}
\author{Takumi \textsc{Nagayama}\altaffilmark{6}}
\author{Ross A. \textsc{Burns}\altaffilmark{7}}
\author{Takeru \textsc{Murase}\altaffilmark{2}}
\author{Ren \textsc{Matsusaka}\altaffilmark{2}}
\author{Makoto \textsc{Nakano}\altaffilmark{8}}
\author{Kazuyoshi \textsc{Sunada}\altaffilmark{6}}
\author{Rin I. \textsc{Yamada}\altaffilmark{9}}
\author{John H. \textsc{Bieging}\altaffilmark{10}}

\altaffiltext{1}{Astronomy Section, Nagoya City Science Museum, 2-17-1 Sakae, Naka-ku, Nagoya, Aichi 460-0008, Japan}
\altaffiltext{2}{Graduate School of Science and Engineering, Kagoshima University, 1-21-35 Korimoto, Kagoshima, Kagoshima 890-0065, Japan}
\altaffiltext{3}{Amanogawa Galaxy Astronomy Research Center (AGARC), Kagoshima University, 1-21-35 Korimoto, Kagoshima, Kagoshima 890-0065, Japan}
\altaffiltext{4}{Centre for Space Research, Potchefstroom campus, North-West University, Potchefstroom 2531, South Africa}
\altaffiltext{5}{Department of Physics and Astronomy, Faculty of Physical Sciences, University of Nigeria, Carver Building, 1 University Road, Nsukka 410001, Nigeria}
\altaffiltext{6}{Mizusawa VLBI Observatory, National Astronomical Observatory of Japan, Osawa 2-21-1, Mitaka-shi, Tokyo 181-8588, Japan}
\altaffiltext{7}{National Astronomical Observatory of Japan (NAOJ), 2-21-1 Osawa, Mitaka, Tokyo 181-8588, Japan}
\altaffiltext{8}{Faculty of Science and Technology, Oita University, 700 Dannoharu, Oita, Oita 870-1192, Japan}
\altaffiltext{9}{Department of Physics, Graduate School of Science, Nagoya University, Furo-cho, Chikusa-ku, Nagoya, Aichi 464-8602, Japan}
\altaffiltext{10}{Steward Observatory, The University of Arizona, Tucson, AZ 85721, USA}

\email{kohno@nagoya-p.jp}
\email{mikito.kohno@gmail.com}

\maketitle

\begin{abstract}
{
We performed NH$_3\ (J,K)=(1,1),~(2,2), {\rm and}~(3,3)$ mapping observations toward the Galactic massive star-forming region Sh 2-255 and Sh 2-257 using the Nobeyama 45-m telescope as a part of the KAGONMA (KAgoshima Galactic Object survey with the Nobeyama 45-metre telescope by Mapping in Ammonia lines) project.
{NH$_3$ (1,1) has an intensity peak {at the cluster S255 N}, is distributed over 3 pc $\times$ 2 pc and is located between two H\,\emissiontype{II} regions.
The kinetic temperature derived from the NH$_3 (2,2)/(1,1)$ ratio was $\sim 35$ K {near the massive cluster S255 IR.}
These clusters also show emission with a large {line width} of $\sim$\,3--4 \kms. Based on the reported data we suggest that \nh\ gas in these regions is affected by stellar feedback from embedded YSO clusters in S255 IR and S255 N.
We also detected NH$_3$ (1,1) emission in a region west of the main gas clump at the location of a concentration of Class II YSOs adjacent to the H\,\emissiontype{II} regions Sh 2-254. 
The presence of Class II YSOs implies $\sim$ 2 Myr of star formation, younger than Sh 2-254 ($\sim 5$ Myr), thus we suggest that star formation in the western region could be influenced by the older H\,\emissiontype{II} region Sh 2-254.}
}
\end{abstract}
\section{Introduction}

\begin{table*}[h]
{
\tbl{Basic parameters of H\,\emissiontype{II} regions and embedded clusters}{
\begin{tabular}{cccccccccc}
\hline
\multicolumn{1}{c}{Name} & $l$  & $b$& log $N_{\rm Lyc}$ &  Exciting Star & Age & References\\
 & [$^\circ$] &[$^\circ$]& [photon s$^{-1}$]   &  & &\\
 (1) & (2) & (3) &(4)& (5) & (6) & (7)  \\
\hline
Sh 2-254 & $192.517$  & $-0.146$ & 47.71 & O9.6 V  & 5.1 Myr &[1,3]\\
Sh 2-255 & $192.630$  & $-0.018$ & 47.92 &  O9.5-B0.0 V & 1.5 Myr &[1,2, 3]\\
Sh 2-256 &  $192.603$  & $-0.129$ & 46.86 &B0.9 V & 0.2 Myr &[1,3]\\
Sh 2-257 &  $192.584$  & $-0.083$ & 47.51  &B0.5 & 1.6 Myr & [1, 2, 3]\\
\hline
S255 IR & $192.600$    & $-0.049$ & 45.11$^*$ & B1$^*$ & $7 \times 10^3$ yr$^{\dag}$ &[2, 4]\\
S255 N &  $192.580$   & $-0.040$ & 45.55  & B1 &$2 \times 10^3$ yr$^{\dag}$  &[2, 4]\\
\hline
\end{tabular}}
\label{param}
\begin{tabnote}
The coordinates of 4 H\,\emissiontype{II} regions indicate the positions of an exciting star \citep{2008ApJ...682..445C}.\\
$\dag$ Dynamical time scale of outflows.
$*$ The value is adopted of the compact H\,\emissiontype{II} region S255-2a. \\
Columns: (1) Name (2) Galactic Longitude (3) Galactic Latitude (4) Number of the Lyman photon (5) Spectral type of the ionizing star (6) Age of the H\,\emissiontype{II} region and embedded cluster (7) References: 
[1] \citet{2008ApJ...682..445C}, [2] \citet{2011ApJ...738..156O}, [3] \citet{2009AJ....138..975B}, [4] \citet{2011AA...527A..32W}\\
\end{tabnote}
}
\end{table*}

Sh 2-255 ($=$S255) and Sh 2-257 ($=$S257) are optically bright H\,\emissiontype{II} regions first cataloged by \citet{1959ApJS....4..257S}. They are confirmed as sites of massive star formation
 belonging to the Gemini OB1 Molecular Cloud Complex (e.g., \cite{1998ApJS..117..387K,1995ApJ...445..246C, 1995ApJ...450..201C, 2017ApJS..230....5W}). 
 {Sh 2-254 ($=$S254) is an older (5.1 Myr) H\,\emissiontype{II} region, while Sh 2-256 ($=$S256) is a younger (0.2 Myr) H\,\emissiontype{II} region, both spatially located close to Sh 2-257 \citep{2008ApJ...682..445C}. This wider region, including multiple H\,\emissiontype{II} regions, is also called the Sh 254-258 ($=$S254-S258) complex.}
These regions have been studied in the optical, infrared, and radio wavelengths for over 4 decades. (e.g., \cite{1974A&A....30..233C,1976A&A....46..153P,1977ApJ...217..448E,1982Ap&SS..87..121M,1983Ap&SS..89..407N,1989ApJ...346..220H,2008ApJ...682..445C,2014MNRAS.439.3719C,2015AA...581A...5S,2015JKAS...48..343L,2021ARep...65..488B, 2021arXiv210612789L}). 

\begin{figure*}
\begin{center} 
\includegraphics[width=18cm]{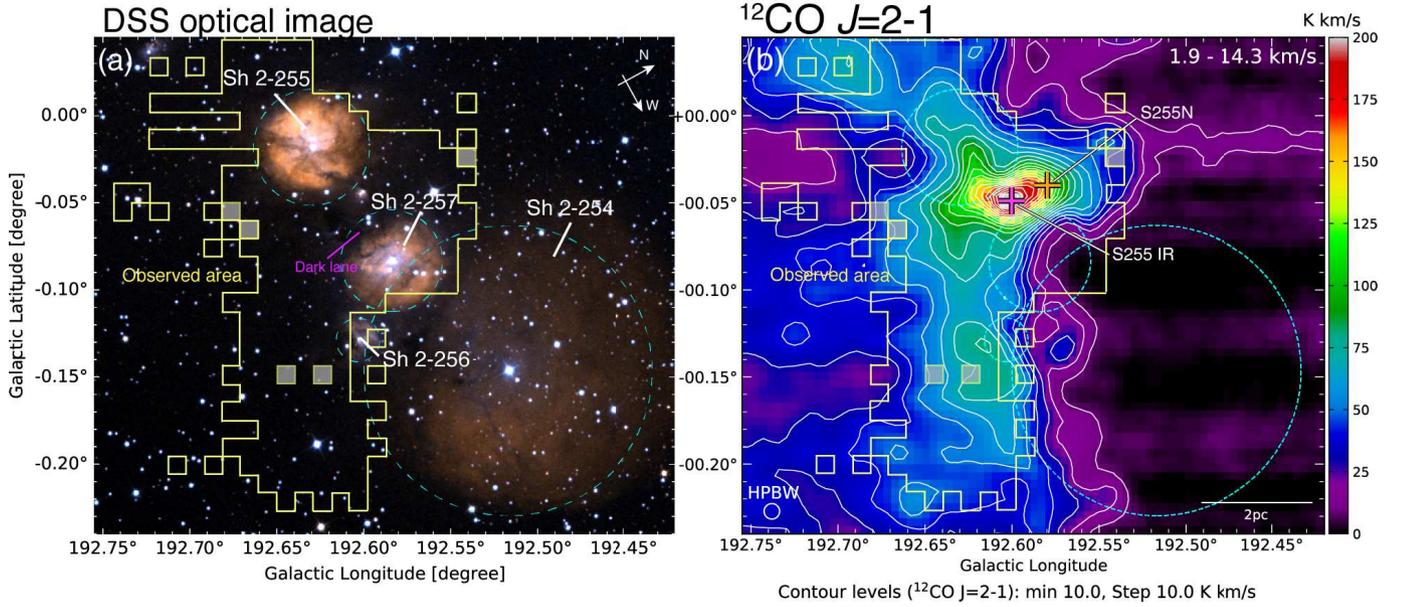}
\end{center}
\caption{(a) Three color composite image of Sh 2-255 and Sh 2-257. Blue, green, and red show DSS2 blue, (DSS2 blue + DSS2 red)/2, and DSS2 red, respectively. (b) $^{12}$CO $J=$2-1 integrated intensity map obtained by \citet{2009AJ....138..975B}. Purple and yellow cross indicates the position of the embedded cluster S255 IR and S255 N, respectively.
The yellow boxes show the \nh\ mapping area of our observation.
{Blue dotted circles enclose the H\,\emissiontype{II} regions. The size of H\,\emissiontype{II} regions are obtained from \citet{1982Ap&SS..87..121M} and \citet{2008ApJ...682..445C}.}}
\label{color}
\end{figure*}

S255 IR ($=$S255-2) and S255 N are located between two H\,\emissiontype{II} regions and house young embedded clusters associated with maser sources (\cite{2011AA...527A..32W}). We summarized the properties of these H\,\emissiontype{II} regions and embedded clusters in Table \ref{param}.
Previous studies suggest that the Sh 254--258 complex, including its H\,\emissiontype{II} regions, are at a common distance of $\sim 2.0$ kpc (e.g., \cite{1995ApJ...450..201C}).
In this paper we adopt a distance of $1.78^{+0.12}_{-0.11}$ kpc obtained by annual parallax measurements of the H$_2$O maser source S255IR-SMA1 \citep{2016MNRAS.460..283B}, noting that this value is consistent with the 1.786 kpc distance estimate for Gem OB1 reported in the Gaia DR2 (see Table 1 in \cite{2019ApJ...879..125Z}).  
Figures \ref{color} (a) and (b) present the Digital Sky Survey (DSS) optical image and $^{12}$CO $J=$2-1 integrated intensity map, respectively, and in which the yellow boxed zone indicates the \nh\ mapping area. 
Sh 2-255 and Sh 2-257 are bright in the optical image, and in their intermediate region exists a dark lane which is seen to obscure the edge of Sh 2-257. 

The $^{12}$CO emission has a peak at the location of S255 IR, which is a massive star-forming region  (\cite{2016MNRAS.460..283B,2020arXiv201208052H,2020ApJ...904..181L}) associated with {6.7 GHz methanol masers (\cite{2010A&A...511A...2R}) and} embedded massive YSOs seen as the three infrared sources of IRS-1 ($=$NIRS-3), IRS-2,  and IRS-3 (e.g., \cite{1997ApJ...481..327H,1997ApJ...488..749M,2001PASJ...53..495I,2015ApJ...810...10Z,2012ApJ...755..177Z,2018ApJ...863L..12L, 2020ApJ...889...43Z,2020PASJ...72....4U}).
S255 NIRS-3 was the first reported case of an accretion burst in a massive protostar (e.g., \cite{2015ATel.8286....1F,2017NatPh..13..276C,2018A&A...612A.103C,2018A&A...617A..80S}).
S255 N, located to the north between the two H\,\emissiontype{II} regions, also shows signs of active star formation and possibly hosts a massive proto-cluster {(\cite{2007AJ....134..346C, 2018ARep...62..326Z})}.

Recently, high-resolution ($\sim 0.1$ pc scale) observations of dense cores in giant molecular clouds were carried out using radio interferometers (e.g., \cite{2019MNRAS.483.3146B}), whereas large-scale \nh\ mapping observations ($\sim 10$ pc scale) aiming to investigate the impact of stellar feedback and interactions in high-mass star formation remain limited. Sh 2-255 and Sh 2-257 are ideal for studying the interaction between H\,\emissiontype{II} regions and molecular clouds.
Therefore, we have carried out NH$_3$ mapping observations toward the Sh 2-255 and Sh 2-257 regions to reveal the dense gas and star-forming mechanism in the giant molecular cloud.

The paper is structured as follows: section 2 introduces observations and archive data sets; section 3 presents the NH$_3$ results; in section 4, we discuss possible star formation scenarios in the Sh 2-255 and Sh 2-257 region, and in section 5 we present a summary of this work.

\section{Observations}
\begin{table*}[h]
\tbl{Observational properties of data sets.}{
\begin{tabular}{cccccccccc}
\hline
\multicolumn{1}{c}{Telescope} & Line& Rest freq.  &Receiver & HPBW  &  Velocity & RMS noise & References \\
& &[GHz] && &Resolution  & level &\\
 (1) & (2) & (3) &(4)& (5) & (6) & (7)  & (8)\\
\hline
Nobeyama 45-m & NH$_3$$(J,K)=(1,1)$& 23.694  &H22 & \timeform{75"}  & 0.39 $\>$km s$^{-1}$ & {$\sim 0.02$} K  & This work\\
 & NH$_3$ $(J,K)=(2,2)$ &23.723 & H22 & \timeform{75"}  &  0.39 $\>$km s$^{-1}$& {$\sim 0.02$ K}  & This work \\
& NH$_3$ $(J,K)=(3,3)$ &23.870 &H22 & \timeform{75"}  &  0.39 $\>$km s$^{-1}$& {$\sim 0.02$ K} & This work \\
\hline
HHT 10-m &$^{12}$CO $J=$ 2--1& 230.538 & & \timeform{32"}  & 1.3 $\>$km s$^{-1}$ & $\sim 0.50$ K  & \citet{2009AJ....138..975B}\\
\hline
\end{tabular}}\label{obs_param}
\begin{tabnote}
Columns: (1) Telescope name (2) Molecular lines (3) Rest frequency (4) Receiver (5) Half-power beam width (6) Velocity resolution (7) r.m.s noise level (8) References
\end{tabnote}
\end{table*}

\subsection{NH$_3$ mapping observations using the Nobeyama 45 m telescope.}

Our observations were carried out as a part of an NH$_3$ survey lead by Kagoshima University using the Kagoshima 6-m \citep{1994vtpp.conf..191O}, Kashima 34-m, and Nobeyama 45-m telescopes (Galactic Centre: \cite{2007PASJ...59..869N,2009PASJ...61.1023N}, NGC 7000: \cite{2011PASJ...63.1259T}, Monkey Head Nebula: \cite{2013ApJ...762...17C}, AFGL 333-Ridge: \cite{2017PASJ...69...16N}, Sh 2-235: \cite{2019PASJ...71...91B}, W33: \cite{2020IAUS..345..353M, murase2021}, Canis Major OB1: \cite{hirata}).

{Mapping observations of NH$_3$ $(J,K)=(1,1), (2,2), (3,3)$, were $J$ and $K$ denote the quantum number of the total angular momentum and projected angular momentum along the molecular axis, respectively \citep{1983ARA&A..21..239H}, were carried out toward the Sh 2-255 and Sh 2-257 massive star-forming region using the Nobeyama 45-m telescope from December 2013 to June 2014, and December 2014 to June 2015 as part of a backup observation program (PI T. Omodaka BU135001 and BU145001).
The rest frequency of NH$_3$(1,1), (2,2), and (3,3) are 23.694 GHz, 23.723 GHz, and 23.870 GHz, respectively.}

Data were acquired using multi-ON-OFF switching observations (three ON points for every one OFF position) toward the Sh 2-255 and Sh 2-257 region adopting a map center position of $(l, b)=(\timeform{192.6285D}, \timeform{-0.0548D})$.
Telescope pointing was adjusted to remain within an accuracy of \timeform{5"} using hourly cross-pattern scans of the H$_2$O maser source S235 $(\alpha_{\rm J2000}, \delta_{\rm J2000}) = (\timeform{05h40m51.96s},\timeform{+35d41'47.0"})$.
The system noise temperature ($T_{\rm sys}$) during observations varied in the range of 120--200 K.
The primary beam half-power beam width (HPBW) and the map grid spacing were \timeform{75"} and \timeform{37.5"}, respectively.
For the system front-end we used the H22 High Electron Mobility Transistor (HEMT) receiver in dual-polarization mode and for the back-end we utilized the digital spectrometer of Spectral Analysis Machine of the 45\,m telescope (SAM45:\cite{Proc..2011,2012PASJ...64...29K}).
 {Eight Intermediate Frequencies (IF) channels were used to record data for the three \nh\ lines and H$_2$O maser simultaneously in dual polarizations. The frequency bandwidth and spectral resolution in each IF were 125 MHz and 30.52 kHz, respectively, corresponding to 1600\,km s$^{-1}$ and 0.39\,km s$^{-1}$ velocity coverage and velocity spacing at 23\,GHz.}

Data were processed using the NEWSTAR\footnote{\url{https://www.nro.nao.ac.jp/~nro45mrt/html/obs/newstar/index.html}} software package \citep{2001ASPC..238..522I}.
We also performed Hanning smoothing (width=5) in the frequency domain to improve the signal-to-noise ratio of detected emission.
 This paper describes the NH$_3$ emission in units of antenna temperature ($T_a^*$) calibrated using the chopper wheel method \citep{1976ApJS...30..247U,1981ApJ...250..341K}.
The root-mean-square noise level for the fully processed data cube after summing the two polarisations was $\sim 0.02$ K.

\subsection{Archival data}
We also make use of $^{12}$CO $J=$2-1 data obtained by the Heinrich Hertz Submillimeter Telescope (HHT: \cite{2009AJ....138..975B}).
We converted the available data cube from the $T_a^*$ scale divided by the main beam efficiency ($\eta_{\rm mb}= 0.75 $) to units of main beam temperature ($T_{\rm mb}$). 
The HPBW and velocity resolution were $\timeform{32"}$ and 1.3 \kms, respectively.
Detailed information regarding the CO observations is presented in \citet{2009AJ....138..975B}, and the basic properties of the \nh\ and CO observations are summarized in Table \ref{obs_param}.

We obtained optical and infrared images from the SkyView web page \citep{1998IAUS..179..465M}\footnote{\url{https://skyview.gsfc.nasa.gov/current/cgi/titlepage.pl}}.
These data are taken from the {Digitized Sky Surveys 2} (DSS2) and the Wide-field Infrared Survey Explorer, WISE telescope \citep{2010AJ....140.1868W}, respectively. 
In addition, we utilize a YSO catalog, including Class I and Class II protostellar objects identified by \citet{2008ApJ...682..445C}.


\section{Results}

\begin{table*}[h]
\tbl{Typical NH$_3$ spectra parameters of the main line}{
\begin{tabular}{cccccccccc}
\hline
\multicolumn{1}{c}{Position}& $l$ & $b$ & Transition &$T_a^*$ & $v_{\rm LSR}$ &  $\Delta v$\\
&[degree]&[degree] & $(J,K)$ & [K] & [km s$^{-1}$] & [km s$^{-1}$] \\
(1) & (2) & (3) &(4)& (5) & (6) & (7)  \\
\hline
A & $192.613$& $-0.045$ &(1,1) & 0.53& 5.8  & 2.1 \\
& &  &(2,2) & 0.24& 5.8  & 2.4 \\
& &  &(3,3) & 0.13& 6.0  & 2.5 \\
\hline
B & $192.572$& $-0.045$ &(1,1) & 0.45& 8.0  & 3.0 \\
& &  &(2,2) & 0.26& 7.8  & 2.9 \\
& &  &(3,3) & 0.14& 7.5  & 3.9 \\
\hline
C & $192.634$& $-0.159$ &(1,1) & 0.27& 7.6  & 1.6 \\
& &  &(2,2) & 0.09 & 7.3  & 1.8 \\
& &  &(3,3) & ---& ---  & --- \\
\hline
\end{tabular}}\label{spec_param}
\begin{tabnote}
Columns: (1) Spectra position (2) Galactic Longitude (3) Galactic Latitude (4) Rotational transition (5) Peak antenna temperature (6) Peak velocity by a single Gaussian fitting (7) The FWHM ($2 \sigma \sqrt{2 \ln 2}$) of spectra by a single Gaussian fitting, where $\sigma$ is the standard deviation of the line profile. 
\end{tabnote}
\end{table*}

\subsection{NH$_3$ spatial distributions}

\begin{figure*}[h]
\begin{center} 
\includegraphics[width=18cm]{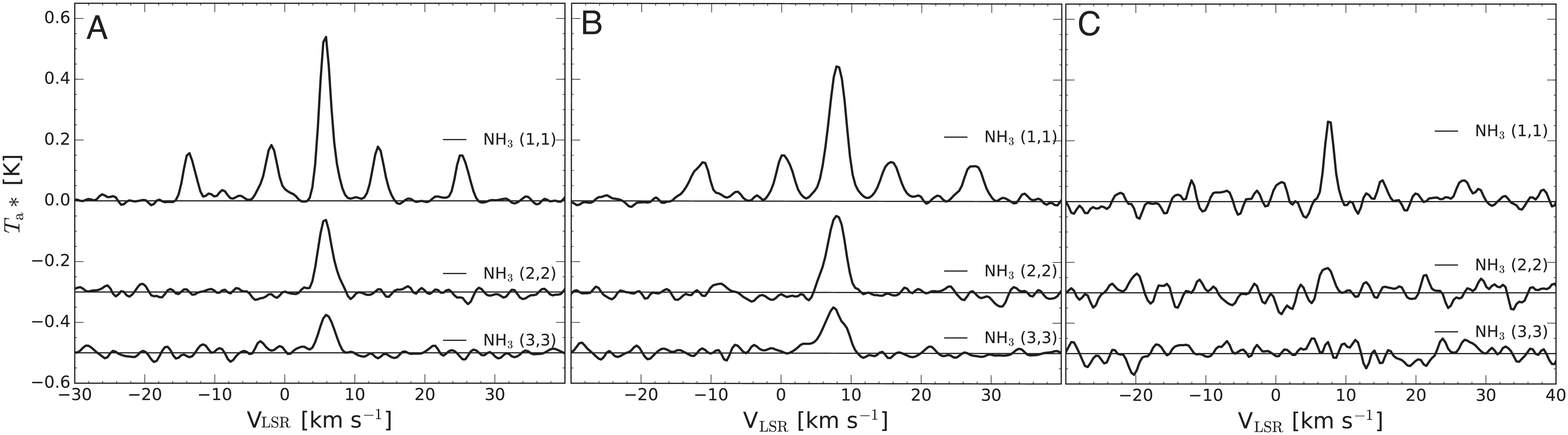}
\end{center}
\caption{NH$_3$ spectra at A $(l,b)=(\timeform{192.613D}, \timeform{-0.045D})$, B $(l,b)=(\timeform{192.572D}, \timeform{-0.045D})$, and C $(l,b)=(\timeform{192.634D}, \timeform{-0.159D})$. These positions are also presented in Figure \ref{integ}(a)}
\label{spec}
\end{figure*}

\begin{figure*}
\begin{center} 
\includegraphics[width=18cm]{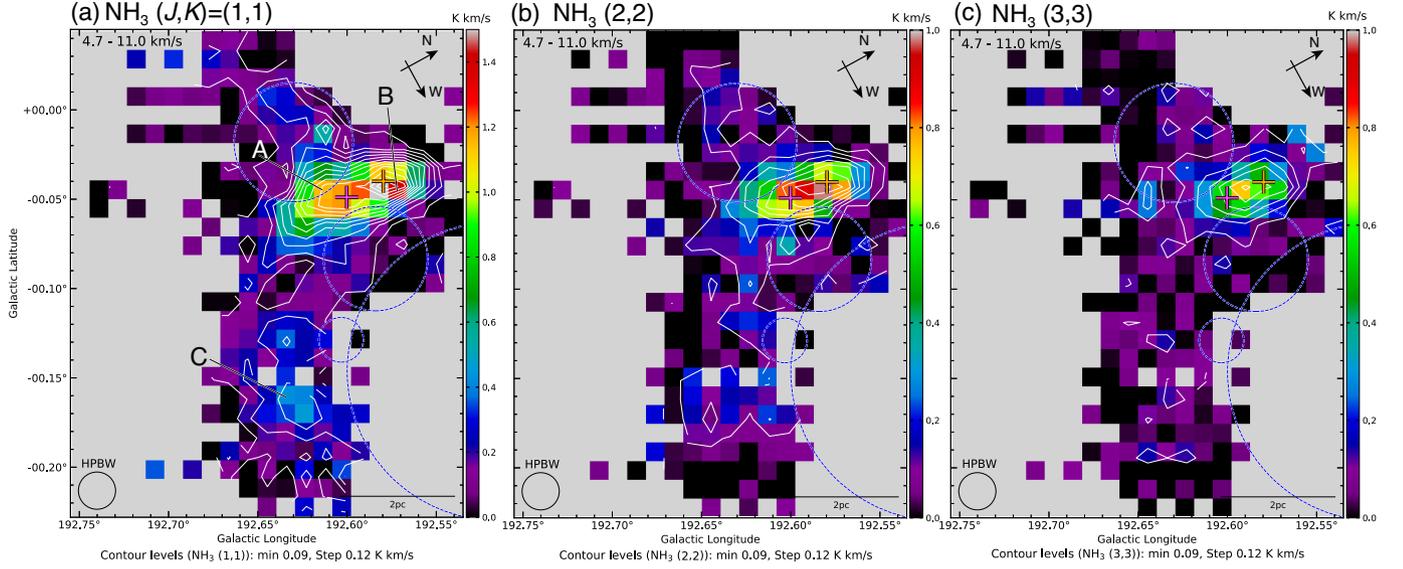}
\end{center}
\caption{Integrated intensity map of (a) NH$_3$ (1,1), (b) NH$_3$ (2,2), and (c) NH$_3$ (3,3). The lowest contour level and intervals are 0.09 K km s$^{-1}$ ($\sim 3\sigma$) and 0.12 K km s$^{-1}$($\sim 4\sigma$), respectively. A, B, and C represent the positions of the spectra in Figure \ref{spec}. The purple, yellow crosses, and blue dotted circles are also the same as Figure \ref{color}(b).}
\label{integ}
\end{figure*}

Figure \ref{spec} (a), (b), and (c) shows the NH$_3$ $(J,K)=(1,1),(2,2)$,and $(3,3)$ spectra at map positions A, B, and C, which signify the locations of S255 IR, S255 N, and a notable western gas clump, respectively. The map locations of these positions are indicated in Figure  \ref{integ} (a).
We detected the inner ($F_1 = 2\rightarrow 1, 1\rightarrow 2$) and outer satellite lines ($F_1 = 0\rightarrow 1,1\rightarrow 0$) of the NH$_3$ (1,1) hyper-fine structure in spectra at positions A and B, where $F_1$ is the quantum number of the total angular momentum including nitrogen spin.
We summarized the properties of each spectrum, obtained by a Gaussian fitting to the spectral data, in Table \ref{spec_param}.


Figure \ref{integ} (a), (b), and (c) show the integrated intensity maps of NH$_3$ (1,1), (2,2), and (3,3), respectively. {The lowest contours represent the detection threshold of $\sim 3\sigma$.}
NH$_3$ (1,1) has a peak at $(l, b)\sim (\timeform{192.58D},\timeform{-0.045D})$ and is distributed over 3 pc $\times$ 2 pc ($l \times b$), elongated in the north-south direction.
This peak is spatially consistent with the embedded cluster S255 N.
On the other hand, NH$_3$ (2,2) and (3,3) emission is seen to peak at $(l, b)\sim (\timeform{192.59D},\timeform{-0.045D})$ and exhibit more compact distributions than NH$_3$(1,1).
The \nh\ clump, including these peaks, exists between two H\,\emissiontype{II} regions Sh 2-255 and Sh 2-257.
Hereafter, we call this dense gas as the ``\nh\ dense clump".
\nh\ spatial distributions are consistent with a previous study presented by \citet{1997A&AS..124..385Z}.
In addition, we also find a weak \nh\ (1,1) peak of the west of the  \nh\ dense clump, at $(l, b)\sim (\timeform{192.63D},\timeform{-0.16D})$.

{

\begin{figure*}
\begin{center} 
\includegraphics[width=16cm]{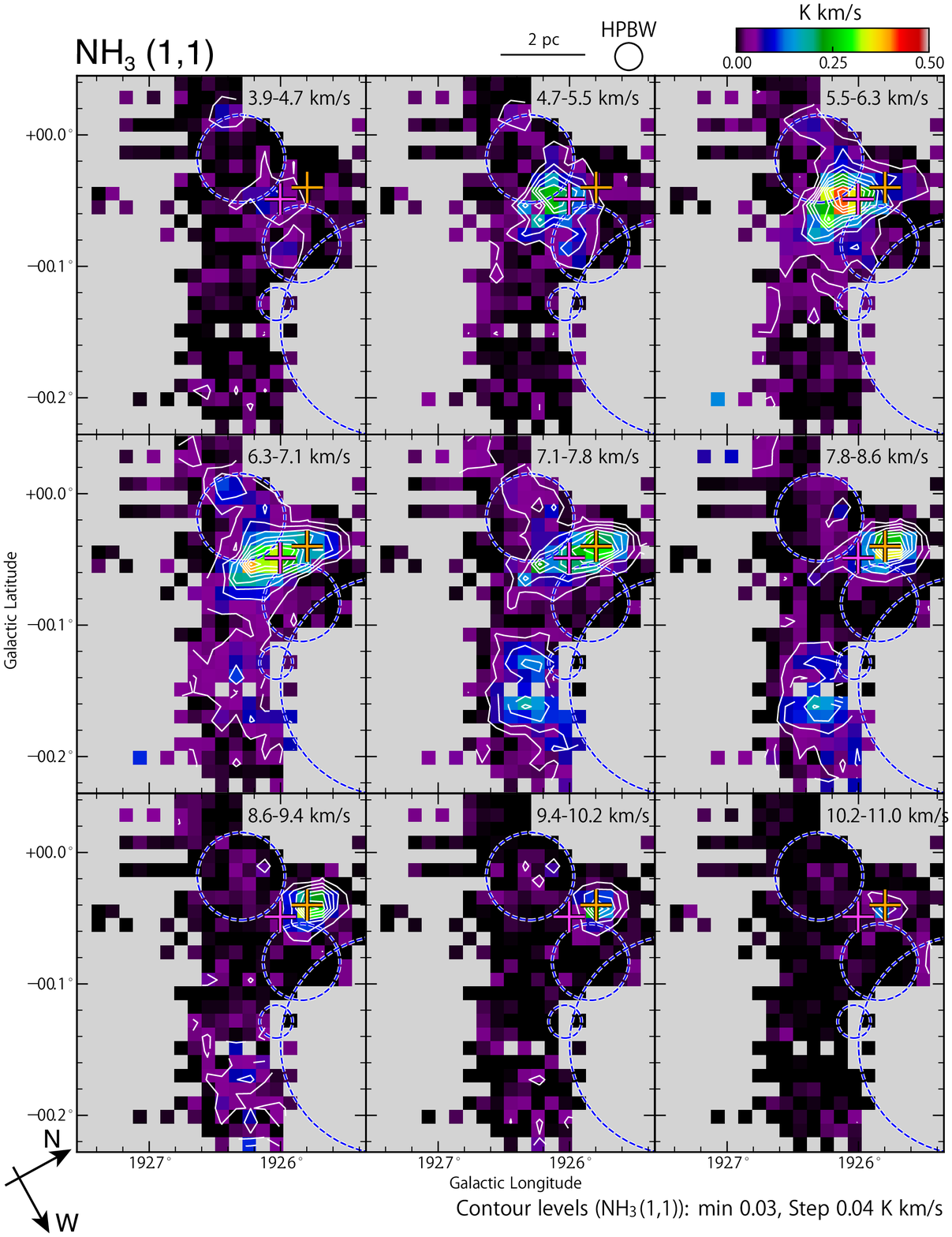}
\end{center}
\caption{{Velocity channel map of NH$_3$ (1,1). The lowest contour level and intervals are 0.03 K km s$^{-1}$ ($\sim 3\sigma$) and 0.04 K km s$^{-1}$ ($\sim 4\sigma$), respectively.  The purple, yellow crosses, and blue dotted circles are also the same as Figure \ref{color}(b).}}
\label{ch}
\end{figure*}

\begin{figure*}
\begin{center} 
\includegraphics[width=16cm]{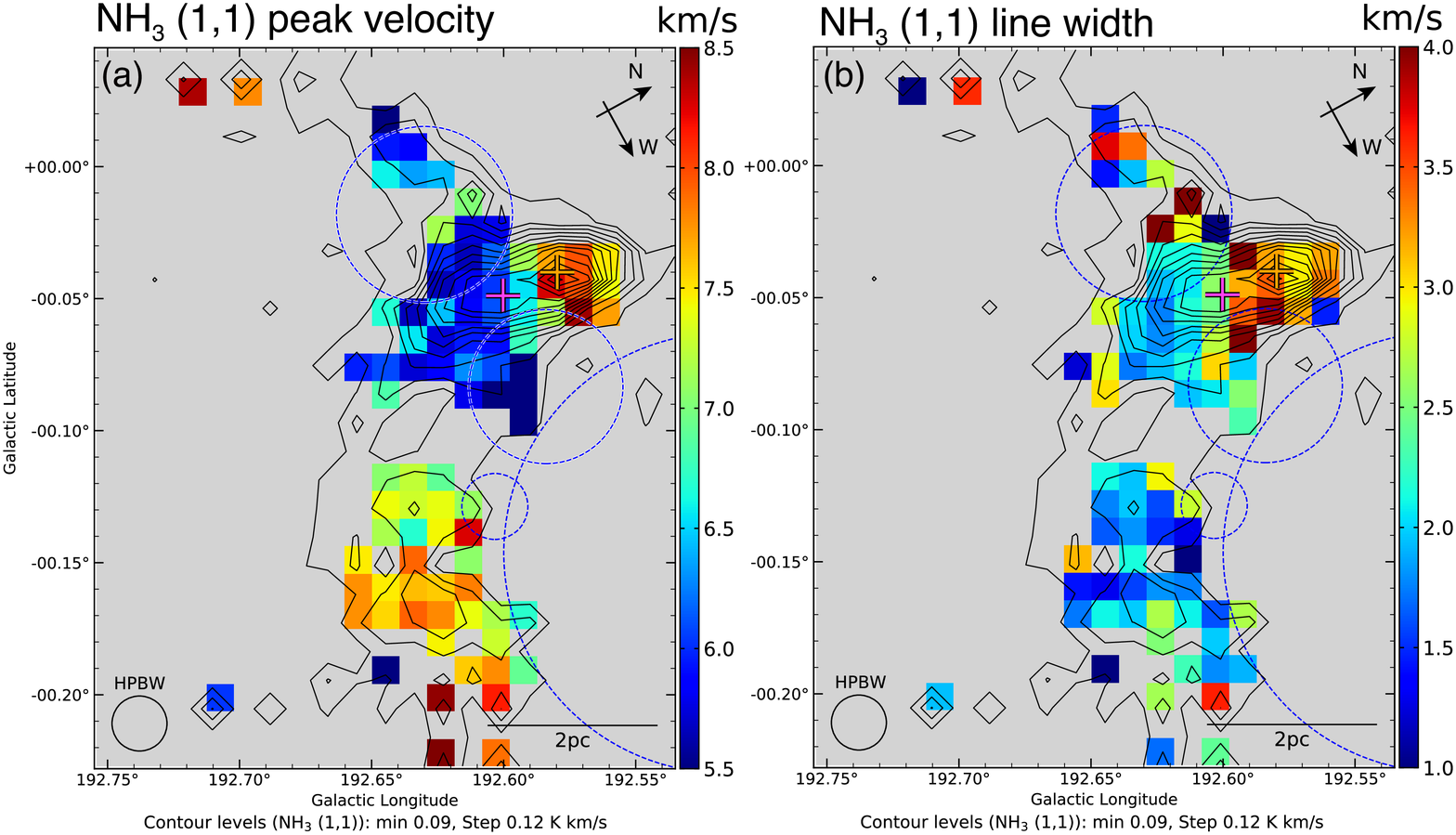}
\end{center}
\caption{{(a) Peak velocity and (b) {line width} maps of NH$_3$ $(J,K)=(1,1)$ by fitting of the Gaussian function adopted of pixels above 0.08 K ($>4 \sigma$). The contours show the NH$_3$ $(J,K)=(1,1)$ integrated intensity. The purple, yellow crosses, and blue dotted circles are also the same as Figure \ref{color}(b).} }
\label{velocity}
\end{figure*}

Figure \ref{ch} presents the \nh\ (1,1) velocity channel maps. 
\nh\ emission in the dense clump corresponding to S255 IR has a central velocity of $\sim 6$ \kms.
The northern intensity peak has a velocity range of 8-10 \kms, which is red-shifted from the center velocity of the \nh\ dense clump.
The weak western emission has a velocity range of 7-8 \kms.
Figure \ref{velocity} (a) and (b) show the peak velocity and {emission line width maps, respectively, in which we derive line intensities and line widths from the height and full width half maximum (FWHM) of a Gaussian fit to the spectral profile at each map position.}


We observed a velocity gradient spanning a range of 5.5-8.5 \kms~ along the direction connecting the northern and southern regions of the \nh\ dense clump (Figure \ref{velocity}a), which corroborates previous results obtained for $^{12}$CO$ J=$3-2 gas by \citet{2009AJ....138..975B} (see their Figure 7, left panel).
The {line width} is widest towards the northern region near S255 IR and S255 N with a width of $\sim 3$-4 km s$^{-1}$, but narrower towards the southern region with a width of $\sim 2$ \kms.
The western peak also has a narrow {line width} of $\sim 1$-$1.5$ km s$^{-1}$ (Figure \ref{velocity}b).

\subsection{Physical parameters of the NH$_3$ dense clump}
We derived estimates of the optical depth, rotational temperature, and total column density using the methods presented by \citet{1983ARA&A..21..239H}, \citet{1992ApJ...388..467M}, and \citet{2015PASP..127..266M}. 
The optical depth can be derived from the ratio of the main and satellite line intensities.
The optical depth $\tau({\rm 1,1,m})$ of the \nh\ (1,1) main line is given by:
\begin{eqnarray}
{T^{*}_{a}({\rm main}) \over T^{*}_{a}({\rm sate}) }={1-e^{-\tau({\rm 1,1,m})} \over 1-e^{-a\tau({\rm 1,1,m})}}
\label{eq:tau}
\end{eqnarray} where $T^{*}_{a}$ is the antenna temperature, and $a$ is the theoretical intensity ratio of the main to the satellite line, which is $a=0.278$ for the main to inner-satellite line ratio, and $a=0.222$ of the main to outer-satellite line ratio \citep{2002ApJ...577..757M}. 
In this paper averaged values of each pair of inner and outer satellite lines were used when deriving ratios to the main line intensity.
To estimate physical parameters we adopted a mean value of optical depth for all map points where emission detections could be made above the $2 \sigma$ noise level.

Assuming the same \nh (1,1) and \nh (2,2) line widths (\cite{2011MNRAS.418.1689U,2018A&A...609A.125W}), the rotational temperature can be derived from the main line intensities of \nh\ (1,1) and \nh\ (2,2) by 
the following equation \citep{1983ARA&A..21..239H} : 


\begin{eqnarray}
T_{\rm rot}(2,2:1,1) &= -41.1 \bigg/ \ln \left[\left({-0.282 \over \tau(1,1,m)}\right)\times \ln \left \{1-{T^{*}_{a}(2,2,m) \over T^{*}_{a}(1,1,m)}\times (1-e^{-\tau(1,1,m)} )\right\}\right]\  [{\rm K}].
\nonumber
\\
\label{eq:trot}
\end{eqnarray}
Figure \ref{trot_column}(a) shows the rotational temperature distribution of \nh\ in the mapped region.
{The center of the \nh\ dense clump near S255 IR has a high rotational temperature of $\sim 25$ K}, while the southern region has a comparatively lower rotational temperature of $\sim 16$ K.
{These values correspond to gas kinetic temperatures of ($T_{\rm kin}$) $\sim 35$ K and $\sim 18$ K, respectively, using a conversion formula from Monte Carlo models (see Appendix B of \cite{2004A&A...416..191T}):
\begin{eqnarray}
T_{\rm kin} &= T_{\rm rot} \bigg/ \left\{1- {T_{\rm rot} \over 41.1} \ln \left[1+ 1.1 \exp \left(-{16 \over T_{\rm rot}} \right) \right]\right\}\  [{\rm K}]
\label{eq:tkin}
\end{eqnarray}}
{The peak gas kinetic temperature is consistent with $T_{\rm kin}$ $= 35$ K obtained by the CH$_3$C$_2$H $J=$ 6-5 data (see Table 8 in \cite{2009MNRAS.395.2234Z}).
We note that the rotational temperature peak is located near S255 IR but does not coincide with it completely (Figure \ref{trot_column}a). The measurement uncertainties might explain this displacement because it is smaller than the HPBW of our observations.}

Using a prescription from \citet{1992ApJ...388..467M}, the NH$_3$ (1,1) column density assuming Local Thermal Equilibrium (LTE) is given by:
}

\begin{eqnarray}
N(1,1) &= 2.78 \times 10^{13} \tau (1,1,m)\left({T_{\rm rot} \over [\rm K]}\right)\left({\Delta v \over [\rm km\ s^{-1}]}\right) [{\rm cm^{-2}}]
\label{eq:n1}
\end{eqnarray}
where $\Delta v$ is the {line width} of the \nh\ (1,1) main line (Figure \ref{velocity}b).

In addition, we estimated the \nh\ total column density of all energy levels 
using $N(1,1)$ (\cite{1991ApJS...76..617T,2013tra..book.....W,2015PASP..127..266M}).
The total column density of \nh\ is derived with the following equation;
\begin{eqnarray}
N_{\rm TOT}(\rm NH_3) &=& {N(J,K) \over g_J \cdot g_I \cdot g_K} Q_{\rm rot} \exp \left({E_u(J,K) \over kT_{\rm rot}}\right) \\
 &\sim& N(1,1) \Biggr[{1 \over 3}\exp \left({23.3 \over T_{\rm rot}}\right) + 1 + {5 \over 3}\exp \left({-41.1 \over T_{\rm rot}}\right) + {14 \over 3}\exp \left({-100.2 \over T_{\rm rot}}\right)\Biggr],
\label{eq:ntot}
\end{eqnarray}
where $Q_{\rm rot}$ is the partition function as follows:
\begin{eqnarray}
Q_{\rm rot} = {\displaystyle \sum_{J}  \sum_{K} } g_J \cdot g_I \cdot g_K  \exp \left(-{E_u(J,K) \over kT_{\rm rot}}\right)
\label{eq:part}
\end{eqnarray}
and in which $k$ is the Boltzmann constant, $E_u(J,K)$ is the energy of the inversion transitions above the ground level, $g_J$ is rotational degeneracy, $g_I$ is nuclear spin degeneracy, and $g_K$ is K-degeneracy. 
Furthermore, we use $E_u (1,1)/k= 23.3, E_u (2,2)/k= 64.4$, and $E_u (3,3)/k= 123.5$ taken from the JPL spectral line catalog\footnote{\url{https://spec.jpl.nasa.gov}} of \citet{1998JQSRT..60..883P}.


\begin{figure*}
\begin{center} 
\includegraphics[width=16cm]{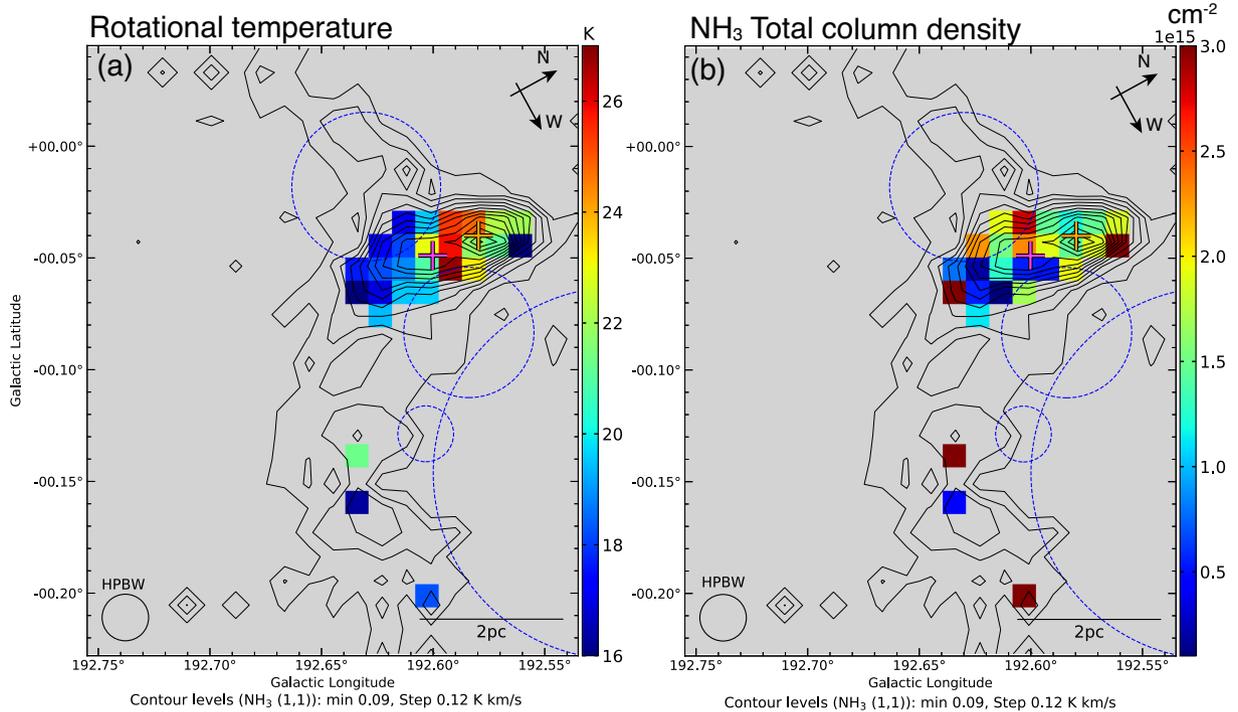}
\end{center}
\caption{{(a) The rotational temperature map obtained by the (2,2)/(1,1) peak intensity ratio above the $4\sigma$ noise level. (b) The NH$_3$ total column density map above the $4\sigma$ noise level. The contours show the NH$_3$ $(J,K)=(1,1)$ integrated intensity.} The purple, yellow crosses, and blue dotted circles are also the same as Figure \ref{color}(b).}
\label{trot_column}
\end{figure*}
Figure \ref{trot_column}\,(b) shows a map of the total column density derived at each observed point.
The column density is $2 \times 10^{15}$ cm$^{-2}$ at the northern region near S255\,N in the \nh\ dense clump while the southern region has a lower column density of $<1 \times 10^{15}$ cm$^{-2}$.
Assuming a H$_2$ conversion factor of $X({\rm NH_3}) = 3.0 \times 10^{-8}$ for the Gemini OB1 \citep{2010ApJ...717.1157D}, 
the column densities of molecular hydrogen in the northern and southern regions were estimated to be $N({\rm H_2}) \sim 7 \times 10^{22}$ and $3 \times 10^{22}$ cm$^{-2}$, respectively. {These values of H$_2$ column density are within the same order of magnitude as the $ \sim 4 \times 10^{22}$ cm$^{-2}$ estimated from CO and 160-500 $\mu$m Herschel data \citep{2021arXiv210612789L}.}

Finally, we estimated the LTE mass from the H$_2$ column density at each map point ($N_i({\rm H_2})$), using the following equation:
\begin{eqnarray}
M_{\rm LTE} &= \mu_{\rm H_2} m_{\rm H} D^2 \sum_{i} \Omega\ N_i(\rm{H_2}),
\label{eq:mass}
\end{eqnarray}
where $\mu_{\rm H_2} \sim 2.8$ is the mean molecular weight contribution of Helium (e.g., Appendix A.1. of \cite{2008A&A...487..993K}), $m_{\rm H}=1.67 \times 10^{-24}$ g is the 
proton mass, $D=1.78$ kpc is the distance to S255 IR, and $\Omega$ is the solid angle. 
We obtained $4.9 \times 10^3$ $M_\odot$ for the LTE mass {which is consistent within an order of magnitude to the value of $\sim 2.6 \times 10^{3}\ M_{\odot}$ obtained for the CO data \citep{2021arXiv210612789L}.  }

\section{Discussion}

\subsection{{Triggered star formation in S255 IR and S255 N dense molecular clumps}}

\begin{figure*}
\begin{center} 
\includegraphics[width=10cm]{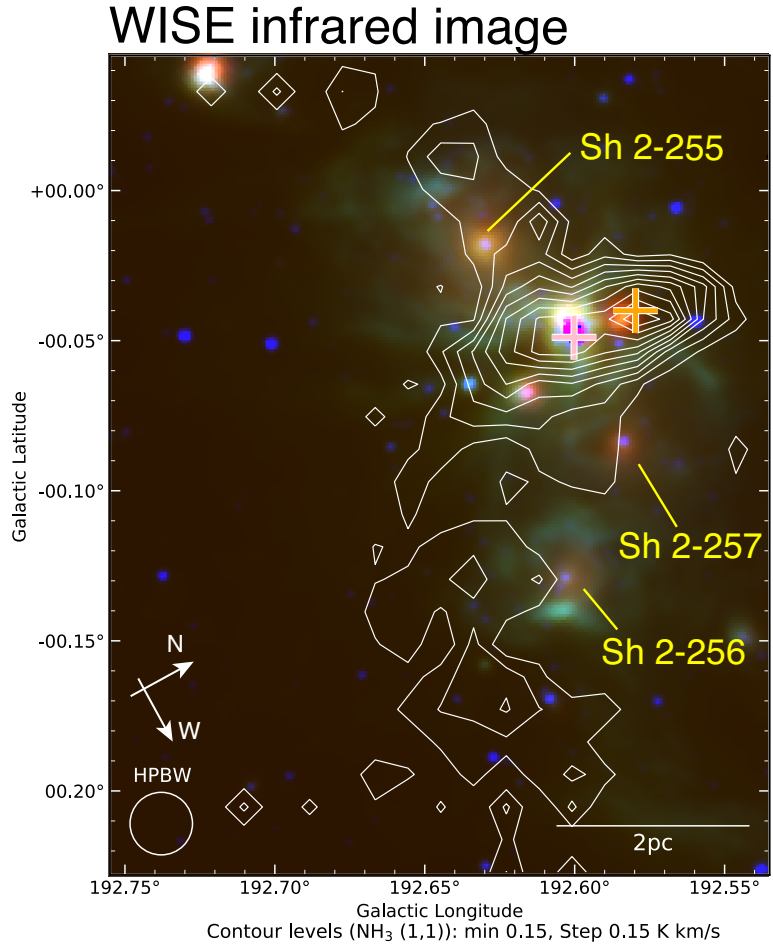}
\end{center}
\caption{NH$_3$ $(J,K)=(1,1)$ distributions integrated with the velocity range from 4.3 to 11.3 km s$^{-1}$ (white contours) superposed on the three color composite infrared image obtained by the WISE satellite. Blue, green, and red indicate $3.6\ \mu$m, $12\ \mu$m and $22\ \mu$m, respectively. The purple and yellow crosses are also the same as Figure \ref{color}(b).}
\label{infra}
\end{figure*}

Figure \ref{infra} shows the \nh\ (1,1) integrated intensity superposed on the WISE infrared image. 
The \nh\ intensity peaks coincide with the infrared peaks of S255 IR and S255 N which have total luminosities of $ 5 \times 10^4\ L_{\odot}$ and $ 1\times 10^5\ L_{\odot}$, respectively \citep{2005A&A...429..945M}.
While \citet{2011AA...527A..32W} pointed out that S255 N is younger than S255 IR, its luminosity is 2 times greater than that of S255 IR.
We also report that the peak of the \nh\ integrated intensity emission is located at S255 N (see Figure \ref{integ}a).
{S255 N resides in a reservoir of dense gas and houses active star formation (e.g., \cite{2007AJ....134..346C, 2018ARep...62..326Z}), which can be expected to produce further generations of new stars.}

Figure \ref{YSO} shows contours of the \nh\ (1,1) integrated intensity superposed on the $^{12}$CO $J=$2-1 integrated intensity map obtained by \citet{2009AJ....138..975B}.
Orange circles represent Class I YSOs, while the white circles indicate Class II YSOs identified by \citet{2008ApJ...682..445C}.
The Class I and Class II YSOs are distributed between Sh 2-255 and Sh 2-257 H\,\emissiontype{II} regions concentrated around S255 IR and S255 N, respectively.
Following \citet{2015ApJ...806..231H}, the time scales of the formation of the Class I and Class II YSOs are $\sim 0.5$ Myr and $\sim 2$ Myr, respectively. 
We therefore infer that S255 IR is in an early stage of star formation based on its concentration of Class I YSOs, identified by \citet{2008ApJ...682..445C}. 

Our observational results also showed large {line width}s and high rotational temperatures around S255 N and S255 IR (Figure \ref{velocity}b and  \ref{trot_column}a).
We suggest that these NH$_3$ dense clumps are heated and experience enhanced turbulent motion locally as a result of feedback from the embedded YSO clusters in S255 N and S255 IR. 
These embedded clusters are likely to be sites of second generation star formation based on their proximity to to Sh 2-255 and Sh 2-257. 
In the case that \nh\ dense clumps are heated by H\,\emissiontype{II} regions directly, the rotational temperatures should be enhanced along the H\,\emissiontype{II} region boundary uniformly because the temperature of ionized gas has $\sim 10^4$ K (e.g., \cite{2011piim.book.....D}).
However, the NH$_3$ gas indicates a low rotational temperature ($ < 18$ K) around $(l,b)\sim (\timeform{192.62D}, \timeform{-0.05D})$ adjacent to Sh 2-255 (see Figure \ref{trot_column}a). 
This result implies that the \nh\ dense clump might not be affected by shock compression due to the expansion of the Sh 2-255 H\,\emissiontype{II} region.
Previous \nh\ mapping observations of other massive star-forming regions reported a similar tendency. For example,
in the Monkey Head nebula, \citet{2013ApJ...762...17C} reported that large {line width}s and high {kinetic temperatures} derived by \nh\ observations are localized around embedded compact H\,\emissiontype{II} regions. 
The authors argued that the dense molecular cloud did not form by the expanding H\,\emissiontype{II} region in the Monkey Head nebula.

{If the Sh 2-255 H\,\emissiontype{II} region is expanding and pushing against the molecular cloud, we would expect to see an arc-like morphology of molecular gas, but such a formation is not seen around Sh 2-255 (Figure \ref{YSO}).
Indeed, \citet{2011ApJ...738..156O} pointed out that the collect and collapse process involving H\,\emissiontype{II} regions might not be effective in the dense clump between Sh 2-255 and Sh 2-257. While the distributions of line width and rotational temperature enhancements of \nh\ gas investigated in this work positionally correlate with the locations of S255 IR and S255 N, our observations of the dense gas traced by \nh\ do not exhibit clear evidence of interaction from the Sh 2-255 H\,\emissiontype{II} region. Consequently, while interaction from Sh 2-255 cannot be ruled out completely, we suggest that other sources of star formation initiation should be considered.}

\subsection{{Impact of the older Sh 2-254 H\,\emissiontype{II} region}}

\begin{figure*}
\begin{center} 
\includegraphics[width=16cm]{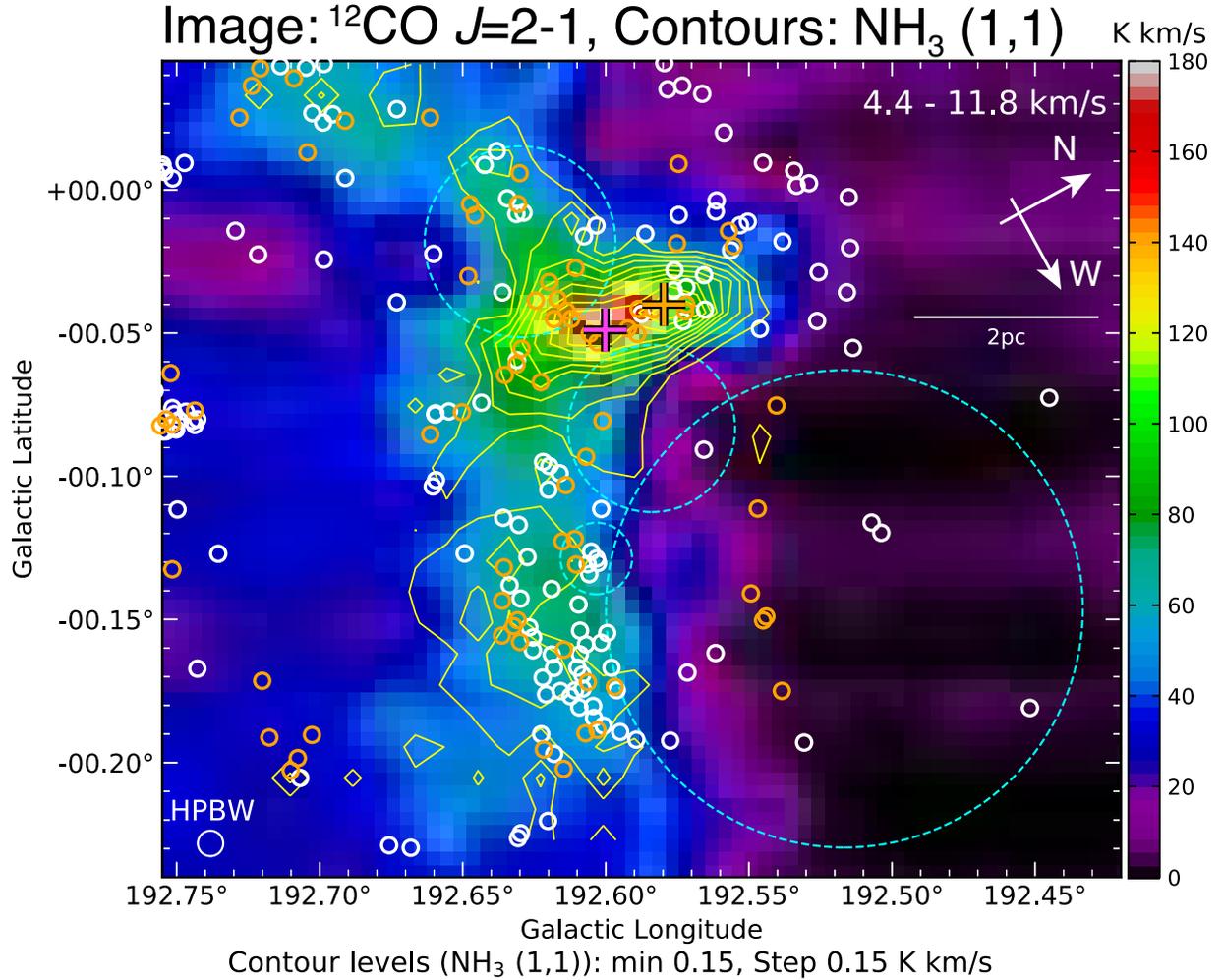}
\end{center}
\caption{NH$_3$ $(J,K)=(1,1)$ distributions integrated with the velocity range from 4.3 to 11.3 km s$^{-1}$ (yellow contours) superposed on the $^{12}$CO $J=$2-1 integrated intensity map with the velocity range from 4.4 to 11.8 km s$^{-1}$\citep{2009AJ....138..975B}. {Orange} and white circles indicate Class I, Class II objects identified by \citet{2008ApJ...682..445C}. The purple, yellow crosses, and blue dotted circles are also the same as Figure \ref{color}(b).}
\label{YSO}
\end{figure*}

Finally, we  discuss the effect of the older H\,\emissiontype{II} region Sh 2-254 which coincides with a hole in the CO emission in this region (see Figure \ref{YSO}).  \citet{2009AJ....138..975B} proposed that molecular gas around Sh 2-254 was ionized by ultraviolet radiation from its exciting star, or swept up by the expanding H\,\emissiontype{II} region.
 We argue that these effects driven by Sh 2-254 are likely responsible for the induced star formation activity around the western NH$_3$ peak. 
The region of western NH$_3$ emission adjacent to Sh 2-254 and Sh 2-256 also contains Class II objects. Therefore, low-mass star formation has progressed for at least $\sim 2$ Myr. This time scale is an order magnitude older than the age of Sh 2-256 ($0.2$ Myr, \cite{2008ApJ...682..445C}).
These Class II low-mass stars were likely to have been formed before triggering by expanding the Sh 2-256 H\,\emissiontype{II} region could have occurred. Consequently, we suggest that the star formation around the western \nh\ might have been caused by the older H\,\emissiontype{II} region Sh 2-254.

The collect and collapse process driven by Sh 2-254 is likely to be responsible for the formation of high-mass protostars (including the exciting sources of Sh 2-256 and Sh 2-257 H\,\emissiontype{II} regions) and low-/intermediate mass stars, particularly the YSOs distributed in the western region of the dense molecular cloud. 
This scenario was suggested previously by \citet{2009AJ....138..975B}, and our results support their hypothesis of triggered star formation in this region.

\section{Summary}
The conclusions of this paper are summarized as follows:
\begin{enumerate}
\item We carried out NH$_3\ (J,K)=(1,1), (2,2), {\rm and}\ (3,3)$ mapping observations toward the Galactic massive star-forming region around Sh 2-255 and Sh 2-257 using the Nobeyama 45-m telescope as a part of the KAGONMA (KAgoshima Galactic Object survey with the Nobeyama 45-metre telescope by Mapping in Ammonia lines) project.
\item NH$_3$ (1,1) gas has a size of 3 pc $\times$ 2 pc (l $\times$ b), an intensity peak at S255 N, and is located between Sh 2-255 and Sh 2-257.
\item The kinetic temperature derived from the NH$_3 (2,2)/(1,1)$ ratio is $\sim 35$ K {near the massive clusters S255 IR.} S255 IR and S255 N exhibit large {line widths} of $\sim 3$-4 \kms. We suggest that \nh\ gas is affected by stellar feedback from the embedded YSO clusters in S255 IR and S255 N.
\item {While the distributions of line width and rotational temperature enhancements of \nh\ gas investigated in this work positionally correlate with the locations of S255 IR and S255 N, our observations of the dense gas traced by \nh\ do not exhibit clear evidence of interaction from the Sh 2-255 H\,\emissiontype{II} region.}
\item {We also detected the NH$_3$ (1,1) emission at a western region adjacent to the H\,\emissiontype{II} regions Sh 2-254 and Sh 2-256. 
This western region has several Class II YSOs with inferred ages of $\sim$ 2 Myr, thus are older than Sh 2-256. 
We conclude that the formation of YSOs at the west side were probably caused by expansion of the older H\,\emissiontype{II} region Sh 2-254.}
\end{enumerate}


\section*{Acknowledgements}
{The authors are grateful to the anonymous referee for their thoughtful comments on the paper.}
The Nobeyama 45-m radio telescope is operated by Nobeyama Radio Observatory, a branch of the National Astronomical Observatory of Japan.

The NH$_3$ back-up observations are promoted on a lot of contributions of Kagoshima university, so the authors would like to thank all them, Dr. Tatsuya Kamezaki, Dr. Gabor Orosz, Dr. Mitsuhiro Matsuo, Mr. Hideo Hamabata, Mr. Tatsuya Baba, Mr. Ikko Hoshihara, and Mr. Masahiro Uesugi.
The authors wish to thank all staff of the Nobeyama radio observatory for helpful backup observations.
We are grateful to Mr. Shun Saeki of Nagoya University for a useful discussion.
The Heinrich Hertz Telescope is operated by the Arizona Radio Observatory, a part of Steward Observatory at The University of Arizona.
The Digitized Sky Survey was produced at the Space Telescope Science Institute under U.S. Government grant NAG W-2166. The images of these surveys are based on photographic data obtained using the Oschin Schmidt Telescope on Palomar Mountain and the UK Schmidt Telescope. The plates were processed into the present compressed digital form with the permission of these institutions.

We acknowledge the use of NASA's SkyView facility (\url{http://skyview.gsfc.nasa.gov}) located at NASA Goddard Space Flight Center.

Software: We used the Astropy, which is the Python package for astronomy \citep{2013A&A...558A..33A, 2018AJ....156..123A}, NumPy \citep{2011CSE....13b..22V}, Matplotlib \citep{2007CSE.....9...90H}, IPython \citep{2007CSE.....9c..21P}, Miriad \citep{1995ASPC...77..433S}, and APLpy \citep{2012ascl.soft08017R}

\end{document}